\documentclass[twocolumn,showpacs,preprintnumbers,amsmath,amssymb,letter]{revtex4}

\usepackage{graphicx}
\usepackage{dcolumn}
\usepackage{bm}

\begin{document}
\title{Experimental evidence for $^{56}$Ni-core breaking from the 
       low-spin structure of the $N=Z$ nucleus $^{58}_{29}$Cu$_{29}$}

\author{A.F.~Lisetskiy,$\,^{1,2}$ N.~Pietralla,$\,^1$ M.~Honma,$\,^3$ A.~Schmidt,$\,^1$
 I.~Schneider,$\,^1$ A.~Gade,$\,^{1,2}$ P.~von Brentano,$\,^1$ T.~Otsuka,$\,^4$
T.~Mizusaki,$\,^5$ and B.~A.~Brown$\,^2$}
\address{ $^1$ Institut f\"ur Kernphysik, Universit\"at zu K\"oln,
               50937 K\"oln, Germany}
\address{$^2$ National Superconducting Cyclotron Laboratory, Michigan State University,
                 East Lansing, Michigan 48824-1321}

\address{$^3$ Center for Mathematical Sciences, University of Aizu, Tsuruga, Ikki-machi,
  Aizu-Wakamatsu, Fukushima 965-8580, Japan}

\address {$^4$ Department of Physics and Center for Nuclear Studies,
               University of Tokyo, Hongo, Tokyo 113-033,
             Japan and RIKEN, Horosawa, Wako-shi, Saitama 351-0198, Japan}

\address{$^5$ Institute of Natural Sciences, Senshu University, Higashimita,
             Tama, Kawasaki, Kanagawa 214-8580, Japan}

\date{\today}

\begin{abstract}
  Low-spin states in the odd-odd $N=Z$ nucleus $^{58}$Cu were
  investigated with the $^{58}$Ni(p,n$\gamma$)$^{58}$Cu fusion
  evaporation reaction at the FN-{\sc tandem} accelerator in Cologne.
  $\gamma\gamma$-coincidences, $\gamma\gamma$-angular correlations,
  and signs of $\gamma$-ray polarizations were measured. 
  Seventeen low spin states below 3.6 MeV
  and 17 new transitions were observed. Ten multipole mixing ratios and
  17 $\gamma$-branching ratios were determined for the first time.
  New detailed spectroscopic information on the $2^+_2$ state,
  the Isobaric Analogue State (IAS) of the $2^+_1,T=1$ state of $^{58}$Ni,
  makes $^{58}$Cu the heaviest odd-odd $N=Z$ nucleus with known
  B(E2;$2^+,T=1 \rightarrow 0^+,T=1$) value. The $4^+$
  state at 2.751 MeV, observed here for the first time, is identified
  as the  IAS of the $4^+_1,T=1$ state in $^{58}$Ni. The new data are
  compared to full $pf$-shell model calculations with the novel
  GXPF1 residual interaction and to calculations within a $pf_{5/2}$
  configurational space with a residual surface delta interaction.
  The role of the $^{56}$Ni core excitations for the low-spin 
  structure in $^{58}$Cu
  is discussed.

\end{abstract}


\maketitle


\section{Introduction}
Odd-odd $N=Z$ nuclei are special many-body systems which are very 
suitable for the test of isospin symmetry \cite{Wilk69,Zu02}. 
The reason is that they are most symmetric 
with respect to the proton-neutron degree of freedom and that yrast 
states with different isospin quantum numbers coexist at low 
energy  \cite{Vog02,Fri99,Lenz99,OLir99,Schm00}.
This allows $\gamma$-ray spectroscopy of 
isovector $(T=1) \to (T=0)$ transitions 
and it makes odd-odd $N=Z$ nuclei important for testing 
isospin symmetry \cite{Zu02,Lis02}. 
Furthermore, these nuclei play a decisive role in the determination 
of the $T=0$ part of effective interactions, {\it e.g.}, 
\cite{Talmi62,Zamick64},  and they are of great interest for the
understanding of weak processes, enhancement mechanisms of
electromagnetic transitions, as well as for problems of nuclear
astrophysics \cite{Town02,Brown01}.

However, up to very recent time more or less comprehensive information
was available only for the odd-odd $N=Z$ nuclei in the $p-$, $sd-$ shells
and for one nucleus from the $pf$-shell: $^{42}$Sc.  Recent progress in both
experimental and theoretical directions brought new valuable data for the
heavy odd-odd $N=Z$ nuclei
$^{46}$V \cite{Fri99,Lenz99,OLir99,Brand01,Brent01,Brent01a,Moel03} ,
$^{50}$Mn \cite{Sve98,Schm00,Lis01,OLear02,Piet02,Hor02},  and
$^{54}$Co \cite{Rud98,Schn00,Zeld99,Brent02} in the lower
part of the $pf$-shell ($f_{7/2}$-shell) and even for some nuclei
of the upper part of the $pf$-shell, like $^{70}$Br \cite{deAn01,Jen02}.
While some understanding of the key problems of the low-energy
structure of $f_{7/2}$ nuclei seems to be obtained and regularities
similar to the ones appropriate for the $sd$-shell are revealed there are
still many uncertainties for low-spin structure 
of these nuclei with mass numbers $A > 56$ 
\cite{Sko98,Rudolph,Vin98,Miz99,Rud99,Mac00,Miz01,Pet01}.
The first
odd-odd $N=Z$ nucleus of this region, which may help to draw confident
conclusions on the situation in the mass region above $^{56}$Ni is 
$^{58}$Cu. But the
experimental data available for the low-energy level scheme of $^{58}$Cu
\cite{Kin71,Bec72,Sta72,Rud73,Rud98a,Fuj98}
is quite sparse and the theoretical full-$pf$ shell model treatment of this
nucleus is a tough computational problem.
Early attempts to understand the low-spin level scheme of $^{58}$Cu 
were, therefore, limited to shell model calculations with
the inert $^{56}$Ni core. 
This approach unsatisfactorily required substantial changes in the
values of single particle energies with respect to the ones for the
$f_{7/2}$ nuclei and too large effective quadrupole charges
\cite{Phill68,bru77}.

This paper presents new experimental data for $^{58}$Cu which was
investigated with the $^{58}$Ni(p,n$\gamma$)$^{58}$Cu fusion
evaporation reaction up to an excitation energy of 3.5~MeV with the
Cologne {\sc Osiris} cube $\gamma$-array.
We could significantly extend
the hitherto known low spin level scheme of $^{58}$Cu
\cite{Kin71,Bec72,Sta72,Rud73}, identify many new transitions, and
establish their multipole character and relative intensities. The new
experimental results are accompanied by full $pf$-shell model
calculations with the new GXPF1 residual interaction \cite{Honma02} universal for
the whole $pf$-shell and help to verify experimental assignments.
The data and the GXPF1 results are
compared to schematic shell model calculations with the $^{56}$Ni
core and a residual surface delta interaction (SDI). The consequences of the softness 
of the $^{56}$Ni core on the spectra are pointed out.

\section{Experimental details and results}
\begin{figure*}
\includegraphics[scale=0.5]{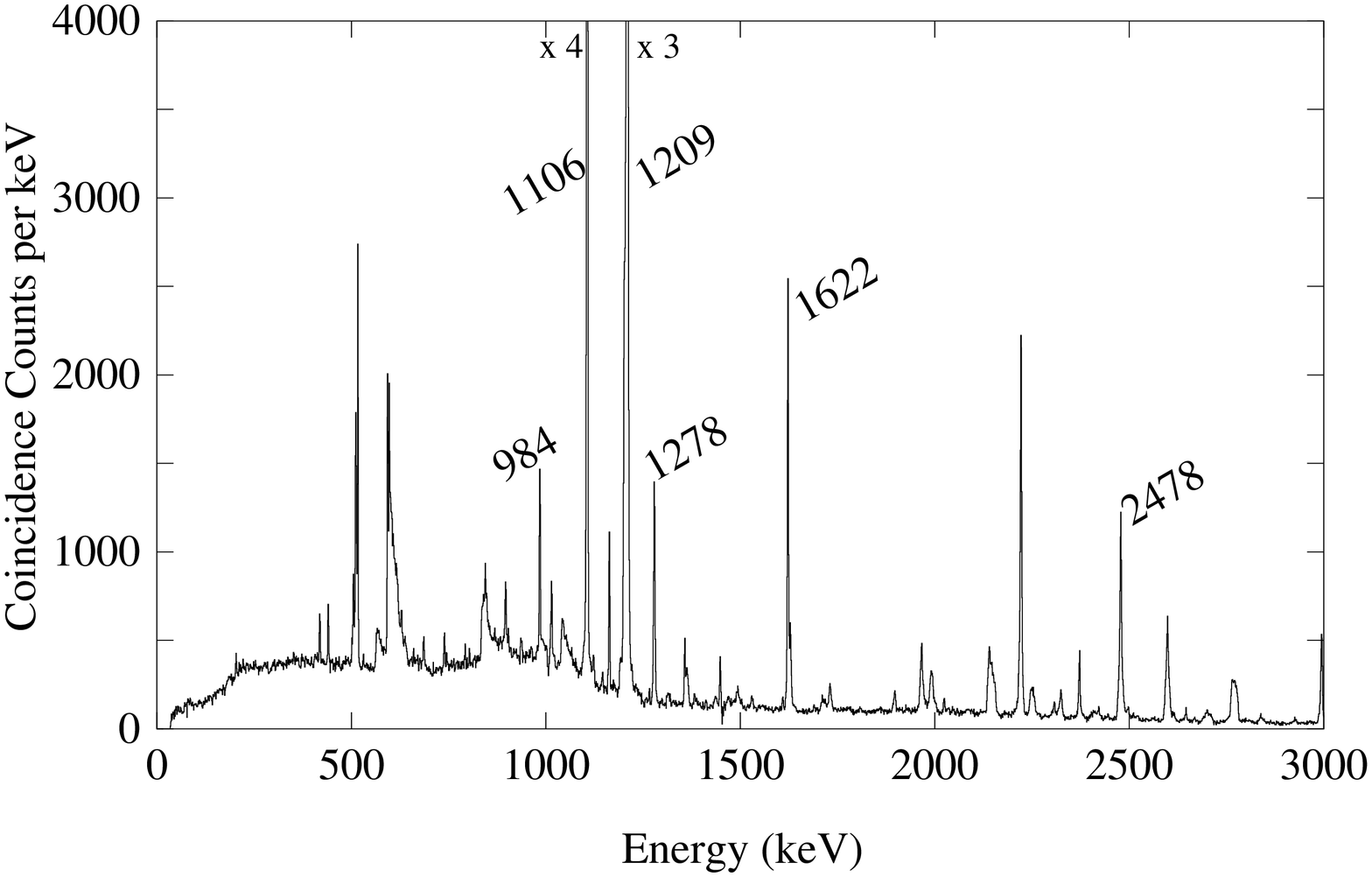}
\caption{The $\gamma$-ray spectrum is obtained by requiring a 
 coincidence condition with the 444~keV $3^+_1\rightarrow 1^+_1$ 
 transition in $^{58}$Cu. The numbers denote energies for 
 transitions between states of $^{58}$Cu (in keV). }
\label{fig1}
\end{figure*}

\begin{figure*}
 \includegraphics[scale=0.65]{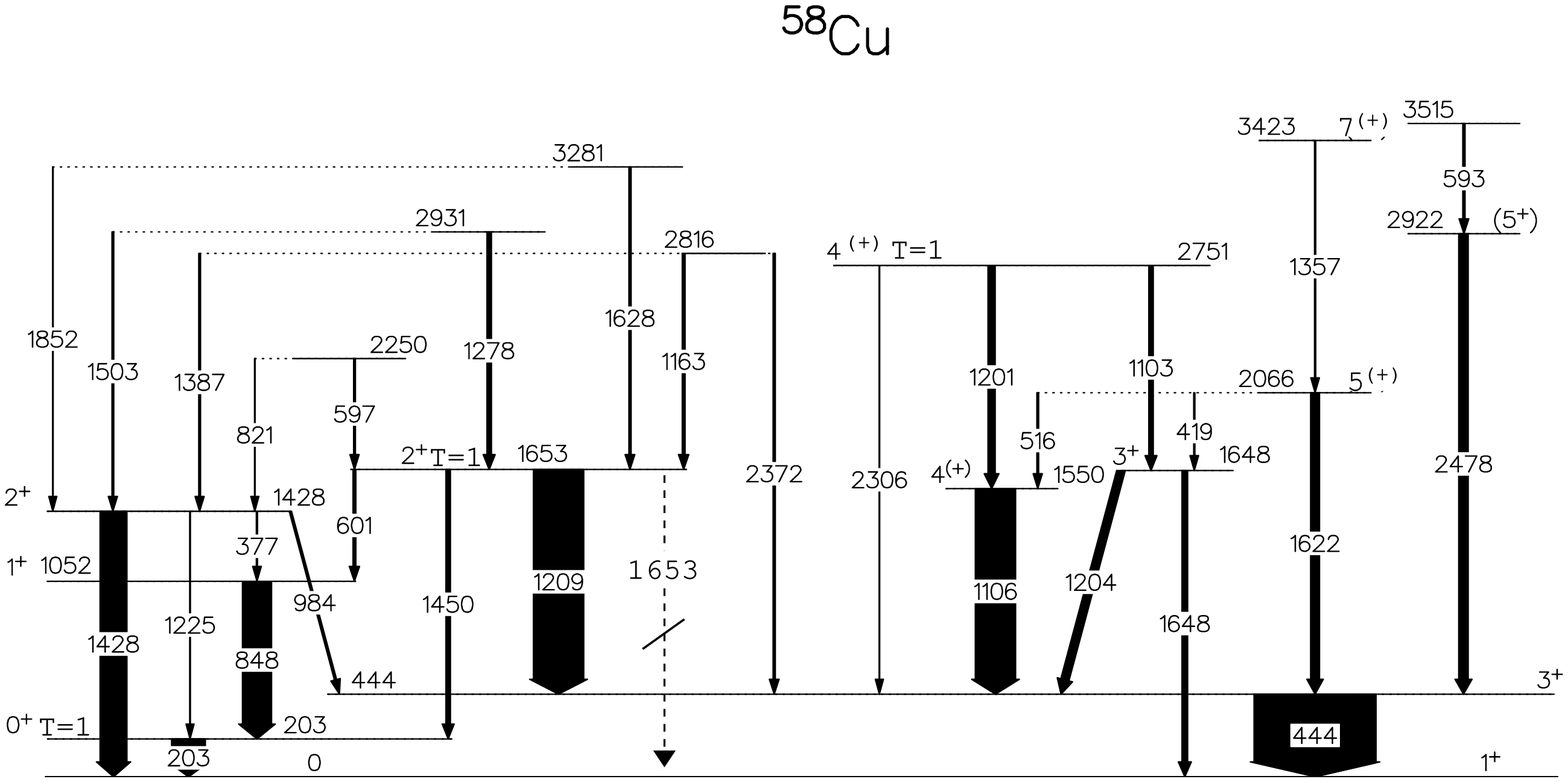}
\caption{Low-spin level scheme of $^{58}$Cu from the $\gamma\gamma$
coincidence relations obtained in the
$^{58}$Ni(p,n$\gamma$)$^{58}$Cu reaction at 14 MeV beam
energy. Levels without an isospin label have $T=0$. 
A possible 1653-keV $2^+_2 \to 1^+_1$ transition marked by the 
dashed arrow has a branching ratio too small to have been detected 
(see discussion).}
\label{fig2}
\vspace*{10mm}

\includegraphics[scale=1.4]{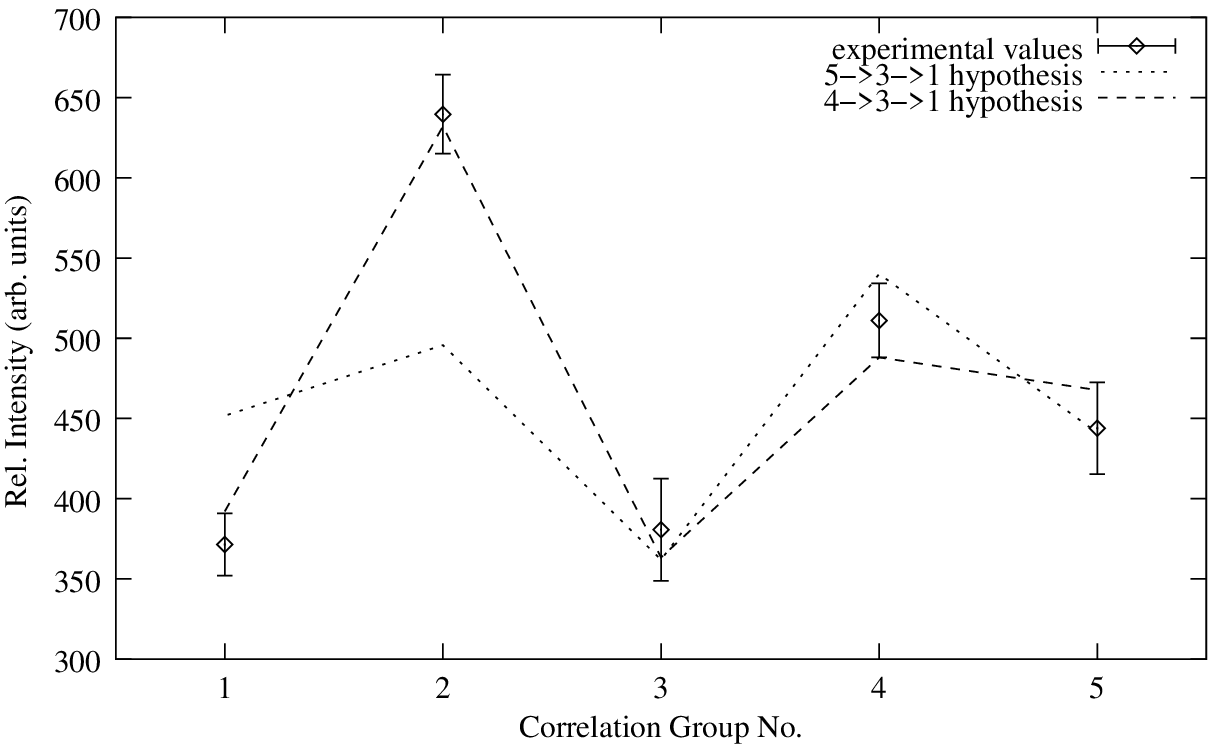}
\caption{Experimental and fitted values of the
$\gamma\gamma$-angular-correlation of the 1103--1648~keV cascade
which connects the $J=4$ level at 2751 keV with the $J^\pi=1^+$ ground
state. Only the $J=4$ spin hypothesis for the upper level at
2751~keV can account for the observed correlation pattern. The fitted
multipole mixing ratio for the $4 \rightarrow 3^+$ transition is
$\delta=-0.07^{+0.05}_{-0.12}$. 
The correlation group nos.~label the different sets of detector 
pairs in the {\sc Osiris} cube spectrometer with common sensitivities 
to the parameters of in-beam $\gamma\gamma$-angular correlation functions 
\protect\cite{Wir}. } 
\label{fig3}
\end{figure*}

Excited states of $^{58}$Cu were populated in the
$^{58}$Ni(p,n$\gamma$)$^{58}$Cu fusion evaporation reaction 
with a beam energy of 14 MeV provided by the Cologne FN-{\sc Tandem}
accelerator. 
The target was a 1 mg/cm$^2$ thick highly enriched self-supporting 
$^{58}$Ni foil. 
Five Compton-suppressed Ge-detectors and one
Compton-suppressed {\sc Euroball Cluster} detectors \cite{Eb96}  
were used in the Cologne {\sc Osiris} cube-spectrometer. 
Two of the Ge-detectors were
mounted in forward direction at an angle $\theta=45^{\circ}$ with
respect to the beam axis. Another two were mounted in the backward
direction at an angle of $\theta=135^{\circ}$ with respect to the beam
axis. The fifth Ge-detector and the {\sc Euroball Cluster} detector were placed
at an angle $\theta=90^{\circ}$ below and above the beam line,
respectively. 
About $10^9$ $\gamma\gamma$-coincidence events were recorded. 
Single $\gamma$ spectra and $\gamma\gamma$ coincidence
spectra of the depopulating photon cascades in $^{58}$Cu were
measured with high energy resolution. As an example of the data,
Fig.~\ref{fig1} shows the $\gamma$ spectrum observed in coincidence
with the decay of the $J^\pi=3^+$, $T=0$ state to the $J^\pi=1^+$,
$T=0$ ground state of $^{58}$Cu.
The low-spin level scheme of $^{58}$Cu was constructed
from the $\gamma\gamma$ coincidence relations.
It is displayed in Fig.~\ref{fig2}. We observed 17
levels and 31 $\gamma$ transitions in this nucleus. With respect to
earlier spectroscopic work \cite{Kin71,Bec72,Sta72,Rud73}, 17 $\gamma$
transitions and five levels are new. In order to assign spin and
parity quantum numbers we analyzed the $\gamma\gamma$ angular
correlation information and the signs of linear polarizations 
using the {\sc Euroball Cluster} detector as a Compton polarimeter. 
The angular correlation pattern
is determined by the spin quantum numbers of the levels involved in a
cascade, by the Gaussian width $\sigma$ of the $m$-substate
distribution of the initial level and by the multipole character of
the corresponding $\gamma$ radiation. 
The Gaussian width $\sigma$ \cite{YAMA} and multipole mixing ratio $\delta$ 
have been deduced from $\chi^2$ minimization \cite{Gad00}.  
The sign convention following Krane, Steffen and Wheeler 
\cite{Kran73} has been used for the determination of $\delta$. 

The analysis of the
$\gamma\gamma$ angular correlations resulted in five new unambiguous
spin assignments for the levels at 444 keV
($J^\pi=3^+$), 1052 keV ($J^\pi = 1^+$),  2066 keV ($J^\pi=5$),  
2751 keV ($J=4$)  and 3423 keV
($J^\pi=7$).  The spin assignments for the levels at 2066 keV 
($J^\pi=5$) and 3423 keV ($J^\pi=7$) are based on the 
spin and parity assignment $J^{\pi}=3^+$ of the level at 444 keV.
The assignment for the level at 1653 keV ($J^\pi = 2^+$)
has been confirmed in the present experiment. 

As an example for the assignments we show in Fig.~\ref{fig3} 
the experimental values of the relative $\gamma\gamma$ 
coincidence intensities of the 1103 -- 1648 keV
cascade for the angular correlation groups of our spectrometer 
together with the values fitted for two different spin
hypotheses. The number of different correlation groups results from
the geometry of the {\sc Cologne-coincidence-cube}-spectrometer
\cite{Wir}. The 1103 -- 1648 keV cascade connects the level at
1648~keV, which could be assigned $J^\pi=3^+$ via the angular
correlation of the 1204 -- 444 keV cascade, with the $J^{\pi}=1^+$
ground state. It is evident from the figure, that a spin quantum number
$J=5$ for the level at 2751~keV can not reproduce the data
($\chi_{min}^2=15.8$) for any value of the possible
octupole/quadrupole mixing ratio $\delta$ of the assumed
$5 \rightarrow 3^+$ transition.
In contrast to this the fitted values are in good accordance with the
experimental ones ($\chi_{min}^2=1.1$) for a spin quantum number $J=4$ for
the level at 2751~keV.   For the correlation analysis we treat the parameter, $\sigma$, 
which describes the Gaussian width of the $m$-substate distribution, as a free parameter.
 Aside from the spin quantum
numbers of the excited states, the measured $\gamma\gamma$ angular
correlations also give valuable information on the multipole mixing
ratios of the $\gamma$ transitions involved (see Table \ref{table1}).

For seven levels with known spin values, we could also deduce the
parity.
This assignment was based on the electric or magnetic character of the
depopulating $\gamma$-transitions.  To determine this character, the
{\sc Cluster} detector was used as a Compton polarimeter. The sum of
two coincident detector signals, which stem from the Compton
scattering of an initial $\gamma$-quantum in one segment of the {\sc
Cluster} and the subsequent absorption in another segment, carries the
full energy information of the initial $\gamma$-ray. The geometry of
the Compton scattering process depends on the polarization of the
initial $\gamma$-ray with respect to the beam axis. Therefore observable
asymmetries of the Compton scattering process allow to measure the
$\gamma$-polarizations and the radiation character.

\begin{figure*}
\includegraphics[scale=0.7]{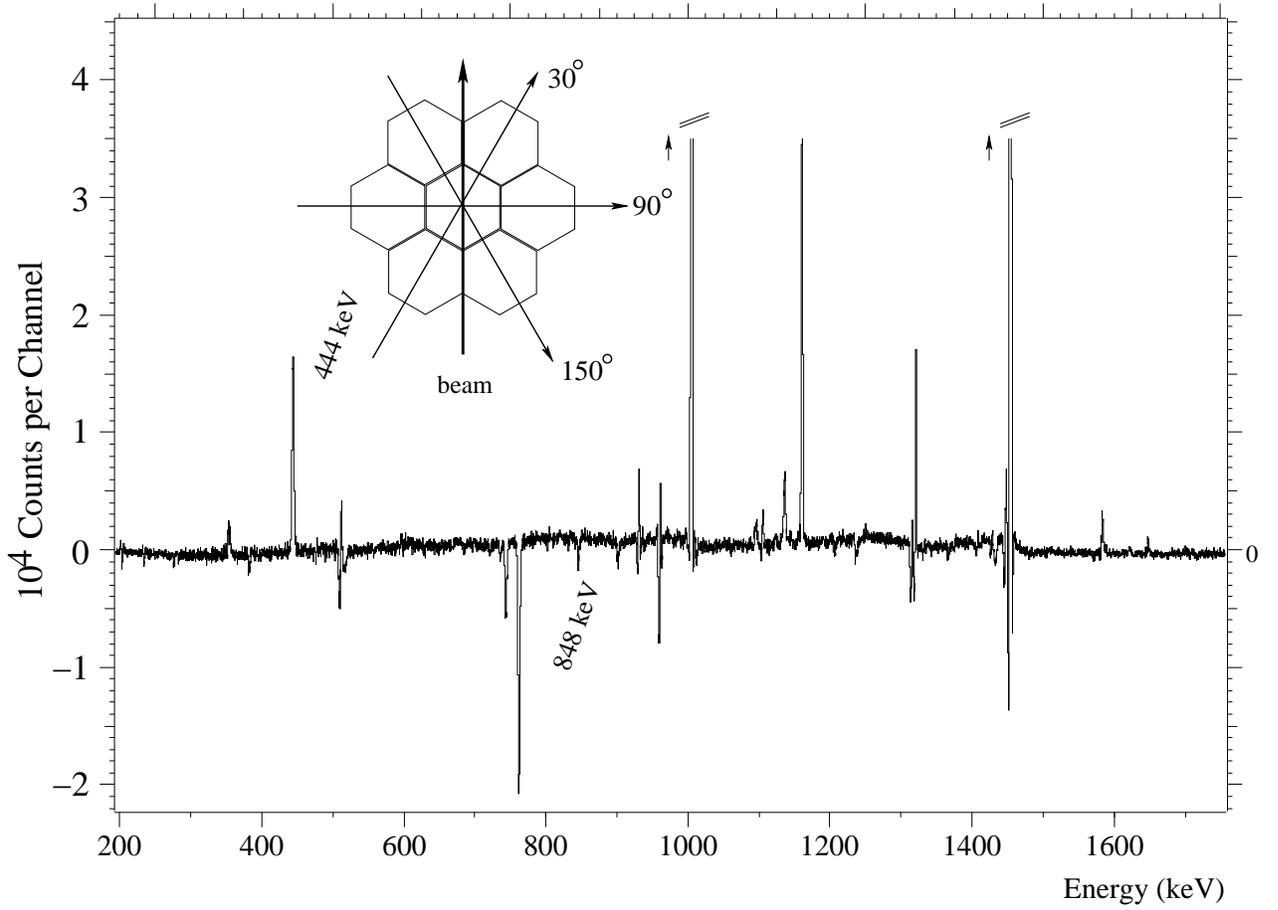}
\caption{Difference spectrum $N_-$ for
initial $\gamma$-rays, which were Compton-scattered and then fully
absorbed in a pair of crystals of the composite {\sc Cluster}
detector with an orientation of $30^\circ$,
$90^\circ$, or $150^\circ$ with respect to the beam axis, as 
is shown in the in-set. One expects
a positive difference $N_{90^\circ}-1/2\,(N_{30^\circ,150^\circ})$
for electric radiation and a negative difference for magnetic
radiation. The largest differences for $\gamma$-lines from
$^{58}$Cu are labeled with the corresponding transition energies.}
\label{fig5}
\end{figure*}

The seven large volume Ge-crystals of the {\sc Cluster} form a
non-orthogonal polarimeter. Numerical simulations \cite{Garc} as well as
recent experiments \cite{Schm00,Schn00,Weiss} have shown, that the
{\sc Cluster}  detector is an efficient Compton polarimeter.  
The in-set in Fig.~\ref{fig5}
shows the configuration of the {\sc Cluster} with respect to the beam
axis in the present experiment. This configuration leads to three
different scattering planes for the Compton scattering of
$\gamma$-rays between adjacent segments of the {\sc Cluster}. In our
experiment these scattering planes enclosed angles of $30^\circ$,
$90^\circ$, and $150^\circ$ with the reaction plane, respectively. The
sum energy of the two coincident signals was sorted in two different
spectra, $N_{90^\circ}$ and $N_{30^\circ,150^\circ}$ depending on to which scattering
plane the involved pair of segments corresponds.
These spectra were used to obtain proper spectra for Compton scattering
intensity differences and sums, namely,
$N_-=N_{90^\circ}-1/2\,N_{30^\circ,150^\circ}$ and
$N_+=N_{90^\circ}+1/2\,N_{30^\circ,150^\circ}$.
The experimental asymmetry is defined as \cite{Weiss}
\begin{equation}
\label{expasym}
A_{\rm exp}=\frac{N_-}{N_+}
\approx Q^{\rm Pol}\,P,
\end{equation}
where $Q^{\rm Pol}$ denotes the positively defined polarization
sensitivity of the {\sc Cluster} and $P$ is the linear polarization of
the incoming photon with respect to the given geometry. Since the
sign of the linear polarization, ${\rm sgn}(P)$, determines the character of
the electromagnetic radiation in case of pure multipolarity, we can conclude
this character with Eq.~(\ref{expasym}) from the sign of the
experimental asymmetry ${\rm sgn}(A_{\rm exp})$.
Fig.~\ref{fig5} shows the difference spectrum $N_-$.
$N_+$ has positive values for all energies.

A summary of the energy levels with certain spin and parity values and
with their depopulating $\gamma$ transitions and branching ratios is
given in Table \ref{table1}.
The assignment of the isospin quantum number $T=1$ is done by
comparing the energies of the levels to the energies of the corresponding
states of the $T=1$ isobaric partner nucleus $^{58}$Ni.

From isospin symmetry we expect, that the excitation energies of
analogue states are close in isobaric  partners.
 The two lowest
excited states in $^{58}$Ni are the $J^\pi=2^+_1$ state at 1454~keV
and the $J^\pi=4^+_1$ state at 2460 keV. From the excitation
energy of the $2^+_1$ state in $^{58}$Ni and from the difference
of excitation energies (1450~keV) of the $0^+_1,T=1$ state
(203~keV) and the $2^+_2$ state (1653~keV) in $^{58}$Cu one can
assign the isospin quantum number $T=1$ to the $2^+_2$ state in
$^{58}$Cu. Furthermore the large $M1$ matrix elements of the 601.4~keV and
1208.8 keV transitions to the $1^+_2$ and $3^+_1$ ,$T=0$ states,
respectively, and the predominantly isovector character of the $M1$
transition operator support the $T=1$ assignment.

Assuming positive parity for the $J=4$ state of $^{58}$Cu at 2751~keV
excitation energy, it can be tentatively
identified as the IAS of the $J^\pi=4^+_1$, $T=1$ state of $^{58}$Ni.
Similar to the case of the $2^+_2$ state,
this assignment is again based on the comparison of the excitation
energy difference (2548~keV) to the $0^+_1,T=1$ state at 203~keV
with the excitation energy of the $J^\pi=4^+_1$ state of $^{58}$Ni 
and on the $\gamma$-decay pattern. 
The decays of that $J^\pi=4$ state to the $3^+_2$ and
$4^{(+)}_1$ $T=0$ states is characterized by very small quadrupole/dipole
mixing ratios.
This fact supports the $T=1$ assignment for the level at 2751~keV.
Although small quadrupole/dipole mixing ratios were also expected for
a $T=1$ state with negative parity we can discard the possibility
of a $T=1$ $4^-$ state because there are no negative parity states in
$^{58}$Ni in this energy region.
In a previous study of $^{58}$Cu \cite{Rud73} a level at 2690(20) keV
was identified as the $T=1$, $J^\pi=4^+$ state from particle spectroscopy in
($^3$He,$t$) charge exchange reactions.
Its $T=1$ assignment was done in Ref.~\cite{Rud73} on the basis of
the close energy match to the $4^+_1$ state of the isobaric partner nucleus
$^{58}$Ni.
This state was not observed in the present experiment and its
decay properties are not known.
Due to the comparably large uncertainty for the excitation energies
deduced from that particle spectroscopy the values of energies from the
previous and the present paper for the assigned $4^+$ states at about 2.7
MeV agree within three standard deviations.
Therefore, one might think that the uncertainty in excitation energy 
of 2690(20) keV
claimed by the authors of Ref.~\cite{Rud73} could have been too optimistic
for that particular level and their $T=1$, $J^\pi=4^+$ state would
coincide with the $T=1$, $J^\pi=4^+$ state at 2751 keV proposed above.
It is, however, also possible that there exists a doublet of $4^+$
states, one with isospin quantum number $T=1$ and the other with $T=0$
as it was recently observed in the neighboring odd-odd $N=Z$ nucleus
$^{54}$Co \cite{Lis02}.
The latter hypothesis is supported by the shell model results discussed
in the next section.

\section{Discussion}
 One of the first successful and very important results of the nuclear
 shell model was an understanding of the origin of the $N=Z=28$ magic
 number. Thus, the nucleus $^{56}$Ni has the properties of a doubly
magic inert core in a simple approach of spherical shell model.
This assumes that the low-energy structure of  $A>56$ nuclei with several 
nucleons above $^{56}$Ni may be described within
 the shell model in small configurational spaces.  $^{58}$Cu is one of
 such nuclei and shell model calculations with a $^{56}$Ni core
have been performed for this nucleus already in the late 1960's
 \cite{Phill68,bru77}.  However, it was realized  that the 
excitations of $^{56}$Ni core are important for the structure of 
the nuclei with $A>56$ \cite{Gou70,Wong68}.  The recent 
information on $^{56}$Ni \cite{Kraus95,Taka98} establishes 
a rather high degree of softness of   $^{56}$Ni.  Core excitations 
are important and can be described by modern large-valence shell 
calculations.  
In fact, it has been found that a recent effective interaction 
suitable for mid-$pf$-shell nuclei produces a significant amount of 
$^{56}$Ni core excitations in neighbor nuclei of about 30 - 40\% 
\cite{Honma02,Honmainprep}. 
One way to identify the impact of the core excitations on the 
structure of $^{58}$Cu is to compare the predictions of modern 
large-scale shell model calculations with an effective interaction adjusted 
for the full pf-shell -- which include the core excitations -- with 
small space shell model calculations with an inert $^{56}$Ni core 
and a schematic interaction suitable for the smaller $pf_{5/2}$ space --  
that do not contain the core excitations. 

\begin{figure*}
\includegraphics[scale=0.6]{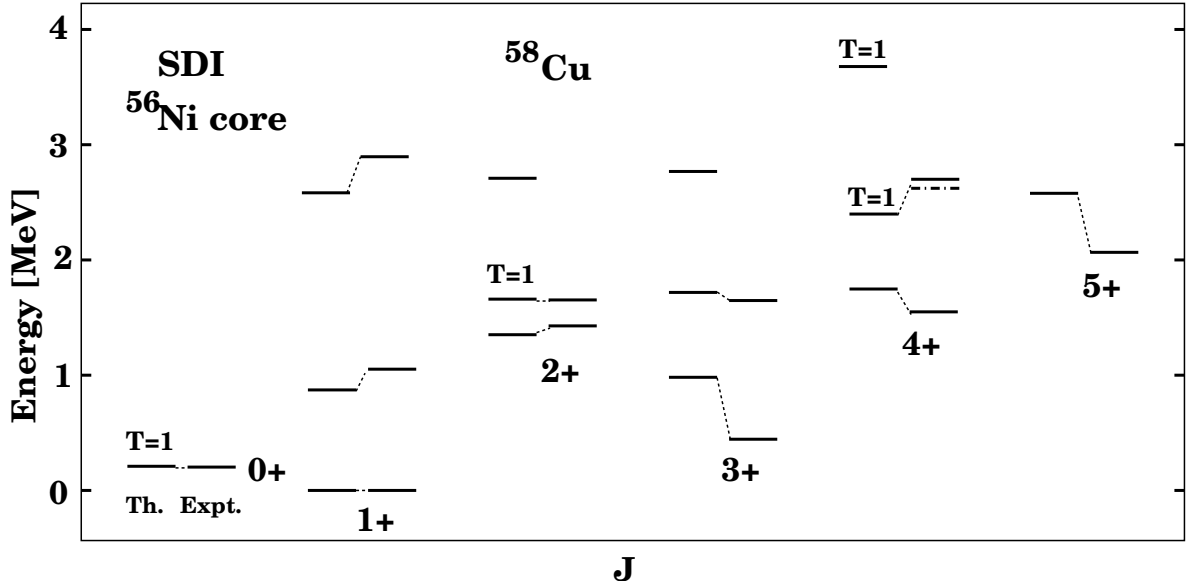}
\caption{Comparison of the calculated (Th.)  and experimental (Expt.)
excitation energies. In the shell model an inert $^{56}$Ni core 
and two-body matrix elements of a residual SDI with
parameterization and single particle energies as shown in
Table~\ref{param} were used.} 
\label{sdispectra}
\end{figure*}

We have, therefore, performed two sets of shell model calculations for
$^{58}$Cu.
The first one uses $^{56}$Ni as the core and a residual surface delta
interaction (SDI) \cite{Mos} with a  parameterization similar to the one
for $^{54}$Co \cite{Lis02}.
Single particle energies were extracted further from the spectrum of $^{57}$Ni
(see Table~\ref{param}). The resulted excitation energies are compared to
 the experimental spectra in Fig.~\ref{sdispectra}. We note that there
 is a good agreement for the two lowest states of each spin value $J$,
 except for the $3^+_1$ and the $5^+_1$ state for which the
 calculated energies are 0.5 MeV higher than the experimental ones.
 We have calculated also $B(M1)$ and $B(E2)$ values between the
 low-lying states which are shown in Table~\ref{4}.

  The second set of calculations was performed by the Tokyo group with the
new effective GXPF1 interaction   \cite{Honma02}.
This interaction
  (195 two-body matrix elements and 4 single particle energies) was determined
partly from a fit to 699 experimental binding energies and level energies from 87 nuclei
   with $A\ge 47$ and $Z\le 32$. The starting point for the fitting procedure
   was a realistic
  G-matrix interaction with core-polarization corrections based
  on the Bonn-C potential.
  Thus, for the first time a universal effective interaction for the whole
  $pf$-shell is determined.

\begin{figure*}
\includegraphics[scale=0.6]{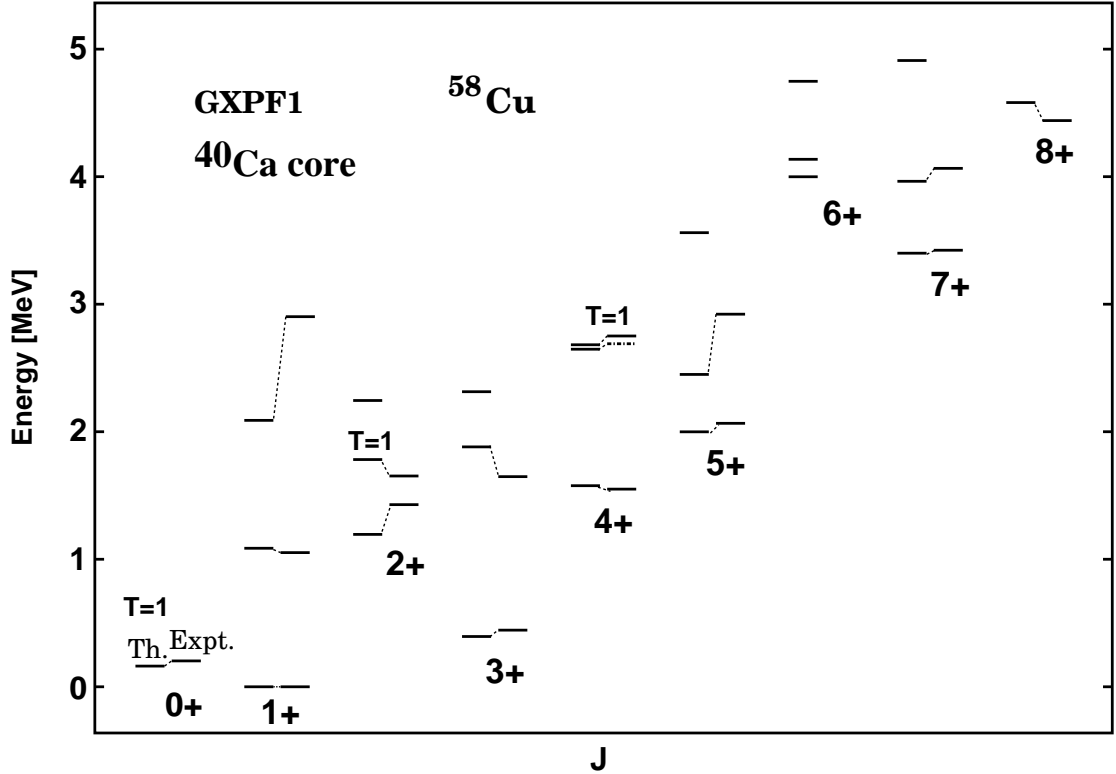}
\caption{Comparison of the calculated (Th.)  and experimental (Expt.)
excitation energies.  In the shell model an inert $^{40}$Ca core 
and two-body matrix elements of the residual GXPF1 interaction
\protect\cite{Honma02} and the corresponding single particle energies (see
Table~\ref{param}) were used.}
\label{gxpf1spectra}
\end{figure*}

  The calculation with the GXPF1 interaction was performed in the
full $pf$ shell with up to 6-particle excitations from the
$f_{7/2}$ orbital to the $p_{3/2}$, $p_{1/2}$ and $f_{5/2}$ orbitals.
Results for the excitation
  energies are compared to the experimental data in Fig.~\ref{gxpf1spectra}.
  One may note quite good reproduction of the experimental data.
  In contrast to the calculations with the inert $^{56}$Ni
  core (see Fig.~\ref{sdispectra}) there are also states with $J>5$ which
  are entirely due to the breaking of the $^{56}$Ni core.  They  are
  coming mainly from a coupling of the one\--neutron\--one\--proton states
  to the first excited $2^+,T=0$ state at 2.7 MeV in the $^{56}$Ni core.
  The energies of the $J^\pi = 7^+_{1,2}$ and $J^\pi = 8^+_{1}$ states are
  perfectly reproduced indicating that the core excitations are correctly
  taken into account. Furthermore there is a much better agreement for the
  $J^\pi = 3^+_{1}$ and the $J^\pi = 5^+_{1}$ states. It has to be emphasized
 also a very good reproduction of the excitation energies and an ordering of
 the $4^+_2,T=0$ and the $4^+_3,T=1$ states which form isospin doublet.
  The electromagnetic transition strengths and lifetimes calculated with
  GXPF1 are compared also to the available experimental data in Table~\ref{4}.

  It is interesting to compare the two sets of calculations.
  The excitation energies of the yrast low-lying states
  with $J \le 5$ are reproduced excellently by the GXPF1 and acceptable
 for the SDI interaction. The mean level deviations are 41 keV and 83 keV, respectively.
  Furthermore
  the single particle energies (s.p.e.) used for the SDI with the $^{56}$Ni
  core and the effective s.p.e. from GXPF1 for the $^{56}$Ni core are rather
  similar. However, switching to electromagnetic transition strengths we
  find many differences (see columns Th-1 and Th-2a, Th-2b of
  Table \ref{4}).

 First, we note that to reproduce the experimental
 $B(E2;2^+,T=1 \rightarrow 0^+,T=1)$ value in the calculations with the
 $^{56}$Ni core we have to increase the sum of the effective quadrupole
 charges $e_p+e_n$ by a factor of 2 as compared to the GXPF1 charges.
 This causes also other $\Delta T = 0$ $E2$ transitions to become
 enhanced some of them even exceeding the  corresponding large $B(E2)$
 values from the GXPF1 calculations, {\it e.g.} like the $E2$ decays of the
 $1^+_2$ state.

  Second, favored isovector $\Delta T = 1$ $M1$ transitions are of
  special interest,  while isoscalar $M1$'s are strongly suppressed and
 usually carry less information on the structure of the wave functions.
 In the simple quasideuteron picture \cite{Lis99} one expects a strong
 enhancement of  $0^+,T=1 \rightarrow 1^+,T=0$ transitions
 (up to 7.3 $\mu_N^2$ with spin-quenching of 0.7) for $^{58}$Cu because of
 the firm presence of the $p_{3/2}$ orbital.
Indeed, in  the calculations with the $^{56}$Ni-core, which closer matches
the quasi-deuteron scheme,
 the summed $B(M1)$ strength for the lowest two $1^+$ states amounts to 5.7
 $\mu_N^2$.
The inclusion of core excitations reduces this sum to 2.5 $\mu_N^2$.
The distribution of this $M1$ strength among these two lowest
 $1^+$ states is different for the two sets of calculations, too.
The $B(M1; 0^+_1 \rightarrow 1^+_1)$ values
 are rather similar in both calculations, but the
$B(M1; 1^+_2 \rightarrow 0^+_1)$ values differ by a factor of 5 even for
the quenching of 0.7 for the SDI.
Since the calculations with the Ni core are in a very small
configurational space one expects larger $B(M1)$ values as compared
  to the large scale shell model.
This is correct for the $B(M1; 1^+_2 \to 0^+_1)$ value but not for
the $B(M1; 0^+_1 \to 1^+_1)$ value.
Furthermore, the comparison of the $E2$ strengths for the
$2^+_1 \rightarrow 1^+_1$ and $2^+_1 \rightarrow 1^+_2$
transitions yields an apparent inversion of the $1^+_1$ and $1^+_2$
  states in the SDI calculation with respect to the GXPF1 results.
The latter yields almost the same $B(E2)$ ratio for these two
transitions as the experimental one.
The $B(E2;3^+_2 \rightarrow 1^+_{1,2})$ values allow to draw the same
conclusion.
This inversion of the $1^+$ states and the considerable reduction of the
$M1$ strengths are caused by the core excitations.

Most interesting are, however, the isovector $M1$ strengths for the
$2^+_2 \rightarrow 1^+_1$ and  $2^+_2 \rightarrow 1^+_2$ transitions.
Their ratio also indicates the inversion of the $T=0$ $1^+$ states:
for the GXPF1 calculations the stronger transition goes to the
$1^+_2$ state, while for the SDI interaction it is the transition to the
$1^+_1$ state.
The latter should be almost completely forbidden according to the 
GXPF1 result.
The $M1$ strengths for the isovector
$2^+_2 \rightarrow 3^+_1$, $2^+_2 \rightarrow 2^+_1$,
$4^+_3 \rightarrow 3^+_{2}$, and $4^+_2 \rightarrow 4^+_{1}$ transitions
indicate that many $B(M1)$ values even from the GXPF1 calculations are
significantly stronger than the isovector  $4^+_3 \rightarrow 3^+_{1}$
or $2^+_2 \rightarrow 1^+_1$ transitions by two to four orders of magnitude.
A suppression of an isovector $M1$ transition by four orders of
magnitude could indicate the presence of a powerful selection rule
being at work.

\begin{figure*}
\includegraphics[scale=0.6]{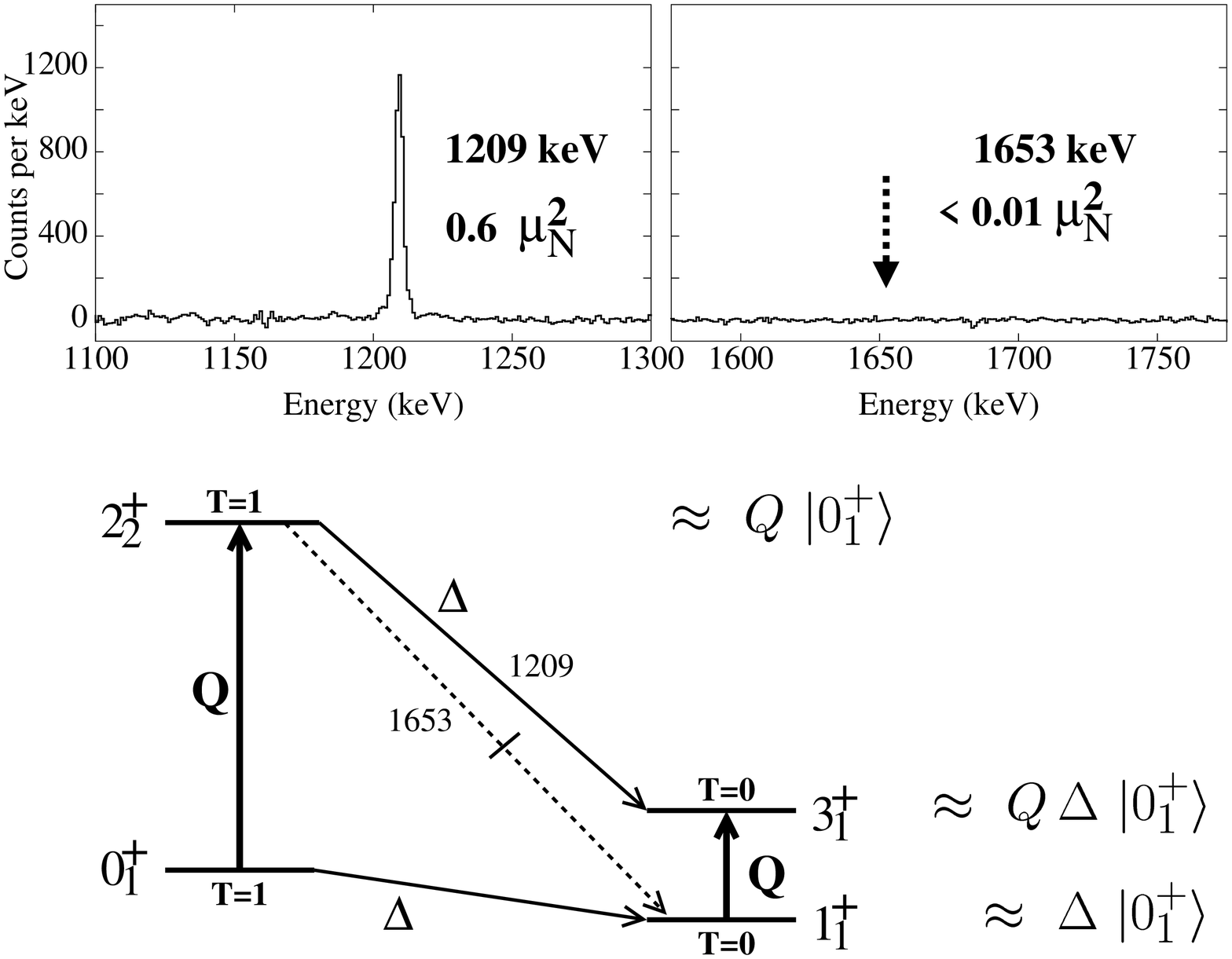}
\caption{Hindrance of the $(T=1) \to (T=0)$ isovector $2^+_2 \to 1^+_1$ 
         $M1$ transition. 
         Relevant parts of the $\gamma$-ray spectra in coincidence 
         with $\gamma$-ray lines feeding 
         directly the $2^+_2$ state of $^{58}$Cu are displayed at the 
         top. 
         A strong $2^+_2 \to 3^+_1$ $\gamma$-ray line is visible at 
         1209 keV while at 
         an energy of 1653 keV no indication for a $2^+_2 \to 1^+_1$ 
         $\gamma$-ray peak 
         was observed in the same spectrum. 
         The intensity branching ratio is $< 4\%$. 
         The $Q$-phonon scheme for $M1$ transitions in the large-scale 
         shell model interprets at the bottom the $2^+_2 \to 1^+_1$ 
         transition as a 
         two-step process and, thereby, explains qualitatively its 
         hindrance.}
\label{fig:Qphon}
\end{figure*}

We propose that this hindrance of the $2^+_2 \rightarrow 1^+_1$ transition
is a consequence of a $Q$-phonon \cite{Sim94,Ots94,Pietr94} selection rule 
applied here to $M1$ transitions in the shell model. 
The reasoning if this interpretation is sketched in Fig.~\ref{fig:Qphon}. 
In the shell model calculation with the GXPF1 interaction
the $T=1$ $2^+_2$ state is most dominantly a complex one-quadrupole
phonon excitation of the $T=1$ $0^+_1$  state, {\it i.e.}, 
to a good approximation $|2^+, T=1\rangle \propto Q\,|0^+, T=1\rangle$ 
where $Q$ denotes the isoscalar quadrupole operator. 
The $T=0$ $1^+_1$ state's wave function is instead generated to a 
large extent from the action of a part $\Delta$ of the isovector 
$M1$ transition operator on the $T=1$ $0^+_1$  state, 
$|1^+, T=0\rangle \propto \Delta\,|0^+, T=1\rangle$ . 
Consequently, the $2^+_2 \to 1^+_1$ transition represents a two-step 
process. 
The one-body $M1$ transition operator cannot simultaneously 
annihilate the $2^+_2 \to 0^+_1$ $Q$-phonon and cause the 
$0^+_1 \to 1^+_1$ $M1$ transition. 
Therefore, the $2^+_2 \rightarrow 1^+_1$ $M1$ transition is strongly
hindered which is well confirmed by the data. 
This interpretation is supported by the strong $2^+_2 \to 3^+_1$, 
$\Delta T =1$ $M1$ transition, which is allowed in the $Q$-phonon 
scheme if we consider the $3^+_1$ state as a $Q$-phonon excitation 
of the $1^+_1$ state. 
For this $2^+_2 \to 3^+_1$ transition one $Q$-phonon excitation is 
present in both, the initial and the final state, and acts as a 
spectator. 
Indeed, the $B(M1)$ values for the $2^+_2 \to 3^+_1$ and 
$0^+_1 \to 1^+_1$ transitions calculated with the GXPF1 interaction 
are close. 
It is of interest to analyze these observations from the viewpoint 
of symmetries discussed in \cite{Isa97,Isa99}.

\section{Conclusion}
In summary we have investigated the low spin states of the odd-odd
$N=Z$ nucleus $^{58}$Cu with the $^{58}$Ni(p,n$\gamma$)$^{58}$Cu
fusion evaporation reaction.
In the present experiment 17 low spin states were observed. 
Five of them and 17 new $\gamma$-ray transitions were observed 
for the first time. 
Numerous multipole mixing ratios and branching ratios  were determined
and 5 new spin assignments were made.
The new data helps to understand the role of core excitations for the
low-spin structure of  $^{58}$Cu.

We have performed shell model calculations for the
low-lying states of $^{58}$Cu with the SDI residual interaction with
a $^{56}$Ni core and with the new GXPF1 interaction which is universal for the
whole pf-shell.  Comparison of the experimental excitation energies to
the corresponding experimental quantities shows that both calculations
yield good agreement for the yrast states with $J \le 5$. However the results
of the two calculations differ considerably for electromagnetic transition
strengths and the agreement with experiment is much better for the full-$pf$ shell
calculations.
In particular, we note that the $B(E2)$ values for isoscalar
transitions are enhanced by a factor of 4 and the isovector $B(M1)$ values are
reduced by factor 5-10 for the full calculation as compared to  the
$pf_{5/2}$ space.  
Big changes in the electromagnetic transition strengths indicate 
the important role of $^{56}$Ni excitations for the structure of 
the low-spin states of  $^{58}$Cu. 
The apparent hindrance of the $(T=1) \to (T=0)$ isovector 
$2^+_2 \to 1^+_1$ $M1$ transition is well reproduced by the GXPF1 
interaction and can be interpreted as the manifestation of a 
$Q$-phonon selection rule for $M1$ transitions in the shell model. 

Another interesting result is the suggested existence of a 
$T=0$ and $T=1$ doublet of
$4^+$ states at $\approx$2.7 MeV. 
The comparison of data with the calculations
favor the $4^+$ state at 2.751 MeV to have isospin $T=1$. 
It would be interesting to find $\gamma$-transitions from the nearby
$4^+,T=0$ state predicted by the shell model and 
ambiguously suggested in \cite{Rud73} at  $\approx$2.69(2) MeV as the
lowest $4^+,T=1$ state of $^{58}$Cu. 
The identification and study of this isospin doublet may offer  valuable
information on the isospin breaking for nuclei along the $N=Z$ line 
above $^{56}$Ni.

\section{acknowledgment}
The authors want to thank in particular A.~Fitzler,
S.~Kasemann, and  H.~Tiesler for help in data taking. We also
thank A.~Dewald, J.~Eberth, A. Gelberg, J.~Jolie,
R.V.~Jolos,  D.~Rudolph,  D.~Wei{\ss}haar, and
K.O.~Zell for helpful discussions.  This work was supported in part by
the DFG under support  No.~Pi 393/1-2, No.~Br -799 /10-2, the U.S. National
Science Foundation Grant No. PHY-0070911 and Grant-in-Aid for Specially
Promoted Research (13002001) from the Ministry of Education,
Science, Sport, Culture and Technology of Japan.  We mention that this work is
originated in a JSPS-DFG joint project. The large-scale numerical calculations were
performed on parallel computers at the Center for Nuclear Study (CNS)
at the University of Tokyo supported by the Grant above mentioned and also by
the CNS-RIKEN joint project for large-scale nuclear structure calculations.

\newpage 
. 
\pagebreak 

\vspace*{10mm}

\begin{table*}[ht]
\caption{Excitation energies $E_i$, spin and parity
quantum numbers $I_i^\pi$ of the initial levels, the measured
$\gamma$-transition energies $E_\gamma$, the excitation energy
$E_f$, and the quantum numbers for the final levels. The last three
columns denote the multipole mixing ratio $\delta$, the
radiation character $\cal{M}\ell$ ($E$ = electric, $M$ = magnetic), and
the relative intensity ratio $I_\gamma$. }
\label{table1}
\begin{tabular}{rcrrcrcr}
$E_i$ & $I^\pi_i$ & $\tau_i$ & $E_\gamma$ & $I^\pi_f$ & $\delta$ & $\cal{M}\ell$ & $I_\gamma$ \\
(keV) & $\hbar$ & (fs)     &  (keV)     & $\hbar$ & & &  \\
\hline
0    & $1^+_1$ & & & & & & \\
\hline
203  & $0^+_1$ & & 203.3 & $1^+_1$ &  & $M1$ & 1\\ \hline
444  & $3^+_1$ & & 444.3 & $1^+_1$ & $-0.02\pm0.04$ & $E2$ & 1\\\hline
1052 & $1^+_2$ & $114(29)$ & 608   & $3^+_1$ &  & $E2$       & $<$ 0.043   \\
     &         &           & 848.8 & $0^+_1$  & & $M1$       & $0.935\pm0.065$ \\
     &         &           & 1052  & $1^+_1$ &  & $E2/M1$ & $<$ 0.087     \\
\hline
1428 & $2^+_1$ & $>$966 & 376.6  & $1^+_2$ & & $E2/M1$ & $0.030\pm0.017$  \\
     & & & 984.2  & $3^+_1$ & -0.84$^{+0.21}_{-1.48}$   & $E2/M1$ & $0.075\pm0.036$  \\
     & & & 1225.1  & $0^+_1$ & 0 & $E2$ & $0.015\pm0.004$  \\
     & & & 1428.3 & $1^+_1$ & & $E2/M1$ & $0.879\pm0.042$  \\\hline
1550 & $4^{(+)}_1$ & $>$505 & 1106.0& $3^+_1$ & $-0.77\pm0.05$ & $(E2/M1)$ & 1\\ \hline
1648 & $3^+_2$ & $>$1312 &  220   &  $2^+_1$ &         & $E2/M1$ & $<$0.034  \\
   &  &    & 1203.5 & $3^+_1$ & $0.53\pm0.13$  & $E2/M1$ & $0.209\pm0.062$ \\
   &  &    & 1647.7 &  $1^+_1$ &  $-0.06^{+0.16}_{-0.27}$  & $E2$ & $0.791\pm0.062$  \\
\hline
1653 & $2^+_2$ & 50(10) & 601.4 & $1^+_2$ & $0.02\pm0.05$  & $M1$ & $0.053\pm0.016^{ }$ \\
& & & 1208.8 & $3^+_1$ & $-0.02\pm0.02$  & $M1$ & $0.900\pm0.027 $ \\
& & & 1449.5  & $0^+_1$ & 0 & $E2$ & $0.048\pm0.016$  \\
 &&&  1653    &  $1^+_1$ &  &  $E2/M1$ & $<$0.037   \\
\hline
2066 & $5^{(+)}_1$ & & 418.6 & $3^+_2$ & & $(M3/E2)$ & $0.073\pm0.036^{ }$\\
& & & 516.3 & $4^{(+)}_1$ & & $(E2/M1)$ & $0.167\pm0.063^{ }$ \\
& & & 1622.0 & $3^+_1$ & -0.12$\pm$0.04  & $(M3/E2)$ & $0.760\pm0.073^{ }$ \\
\hline
2250 & & & 596.7  & $2^+_2$ & &  & $0.894\pm0.048$ \\
& & & 821.3  & $2^+_1$ & &  & $0.106\pm0.048$  \\
\hline
2751  & $4^{(+)}$ & & 1103.1  & $3^+_2$ & $-0.07^{+0.05}_{-0.12}$  & $(E2/M1)$ & $0.397\pm0.081$ \\
& & & 1200.6  & $4^{(+)}_1$ & $0.00\pm0.05$  & $(M1)$ & $0.554\pm0.084$  \\
& & & 2306.4  & $3^+_1$ & & $(E2/M1)$ & $0.049\pm0.018$  \\ \hline
2816 & & & 1162.7  & $2^+_2$ & & & $0.419\pm0.082$ \\
& & & 1387.2  & $2^+_1$ & &  & $0.241\pm0.057$  \\
& & & 2371.5  & $3^+_1$ & &  & $0.340\pm0.076$  \\
\hline
2922 & $(5^+) $ & &  856    & $5^{(+)}_1$ & & $(E2/M1)$ & $<$ 0.028  \\
     &           & & 1274    & $3^+_2$     & & $(E2)$    & $<$ 0.028  \\
     &           & & 1372    & $4^+_1$     & & $(E2/M1)$ & $<$ 0.028  \\
     &           & & 2477.5  & $3^+_1$     & & $(E2)$    & $0.959\pm0.041$\\

\hline
2931  & & & 1278.3  & $2^+_2$ & & & $0.765\pm0.062$  \\
& & & 1503.0  & $2^+_1$ & &  & $0.235\pm0.062$  \\
\hline
3281 & & & 1627.7  & $2^+_2$ & & & $0.843\pm0.053$  \\
& & & 1852.2  & $2^+_1$ & &  & $0.157\pm0.053$ \\
\hline
3423 & $7^{(+)}$ & & 1356.7 & $5^{(+)}_1$ & & $(M3/E2)$ & 1\\
\hline
3515 & & & 592.7 & $5^{(+)}_1$ & & & 1\\
\end{tabular}
\end{table*}

\begin{table*}
\caption{The interaction parameter of the Surface Delta Interaction as
defined in \protect\cite{bru77}, the single particle energies (s.p.e.) of the orbits
included, effective $e_p$ and $e_n$ charges, effective g-factors for SDI and GXPF1 as well as  effective s.p.e.
for GXPF1 \protect\cite{Honma02}. Although, those are not parameters we show effective
single-particle energies for the GXPF1 interaction in the column of the s.p.e.}
\label{param}
\begin{center}
\begin{tabular}{c|cccc|ccc|cc|cccc}
Int. & \multicolumn{4}{c|}{ s.p.e. (MeV)} &  \multicolumn{3}{c|}{ Parameter values (MeV)} & \multicolumn{2}{c|}{eff. charges}  & \multicolumn{4}{c}{eff. g-factors}\\ \hline
     & $\varepsilon_{f_{7/2}}$     &
       $\varepsilon_{ p_{3/2}}$    &
       $\varepsilon_{\nu f_{5/2}}$ &
       $\varepsilon_{\nu p_{1/2}}$ &    A$_{T=1}^{\rho \rho}$ & A$_{T=0}^{pn}$ & B  & $e_p$ & $e_n$ & $g_l^p$ & $g_l^n$ & $g_s^p$ & $g_s^n$\\
\hline
SDI  (Th-2a)  & -     & 0.00 & 0.83 & 1.88  & 0.50 & 0.45 & 0.16 & 1.50 & 0.50 & 1.00 & 0.00 & 5.59 & -3.83 \\ \hline
SDI  (Th-2b)  & -     & 0.00 & 0.83 & 1.88  & 0.50 & 0.45 & 0.16 & 2.50 & 1.50 & 1.00 & 0.00 & 3.91 & -2.68 \\ \hline
GXPF1 (Th-1)
       & -7.00 & 0.00 & 1.00 & 2.50 &  & & & 1.5  & 0.5  & 1.00 & 0.00 & 5.59 & -3.83 \\ \hline
\end{tabular}
\end{center}
\end{table*}
\begin{table*}
\caption{Calculated and experimental electromagnetic transition strengths and
lifetimes in $^{58}$Cu. The experimental energies were used for
calculations of lifetimes. The results are shown for the GXPF1 interaction
(Th-1) and for the SDI (Th-2a and Th-2b). Free g-factors
$g_s^{\rm eff}=1.0\cdot g_s^{\rm free}$ and effective quadrupole charges
$e_p=1.5$, $e_n=0.5$ were used for Th-1 and Th-2a while
$g_s^{\rm eff}=0.7\cdot g_s^{\rm free}$ and $e_p=2.5$, $e_n=1.5$ for the Th-2b.
The B(M1) values smaller than $10^{-4}$ are replaced by 0.0. The quantities ``x'' and ``y'' are
introduced for the $5^+_1$ and the $4^+_3$ states, respectively,  in order  to show
experimental ratios of corresponding B(E2) or B(M1) values.}
\label{4}
\begin{center}
\begin{tabular}{|cc|ccc|cccc|cccc|cc|}
\hline
$J_i,T_i$ & $J_f,T_f$ &
\multicolumn{3}{c|}{$E_i$ (MeV)} &
\multicolumn{4}{c|}{B(E2;$J_i\rightarrow J_f$),$[e^2$fm$^4]$} &
\multicolumn{4}{c|}{B(M1;$J_i\rightarrow J_f$),$[\mu_N^2]$} &
\multicolumn{2}{c|}{Lifetime, $\tau_i$} \\
\hline
 & & Expt. & Th-1 &  Th-2 &
     Expt. & Th-1  &  Th-2a  &  Th-2b &
     Expt. & Th-1 &  Th-2a &   Th-2b &
     Expt. & Th-1. \\
\hline
$0_1^+,1$ & $1_1^+,0$ & 0.203 & 0.162 & 0.210 & & & & &  & 1.58 & 2.32  & 1.05 &  & 4.3 ps \\
\hline
$3_1^+,0$ & $1_1^+,0$ & 0.444 & 0.394 & 0.980 & & 84 & 2 & 9 & & & &  & & 0.56 ns \\
 \hline
$1_2^+,0$ & $1_1^+,0$ & 1.051 & 1.086 & 0.872 & $<$ 695 & 37 & 32 & 129 & $<$ 0.054 & 0.01&  0.01 & 0.001 & 114(29) fs & 287 fs \\
          & $0_1^+,1$ &  & &  & & & & & 0.78(24) & 0.30 & 2.94  & 1.55 & & \\
          & $3_1^+,0$ &  & &  & $<$ 5329 & 30& 56 & 224 & & & & & & \\
\hline
$2_1^+,0$ & $1_1^+,0$ & 1.428 & 1.195 & 1.351 & $b \times 3.2$\footnote{$b<41$} & 3.7 & 40 & 162 & $<$ 0.02 & 0.005 & 0.01 & 0.001 & $>$ 1.0 ps & 2.6 ps \\
          & $0_1^+,1$ & & &   & $b \times 0.15$  & 0.21 & 0.3 & 0.3 & & & & & & \\
          & $3_1^+,0$ & & &   & $b \times 2.1$  & 54.6 & 12 & 48 & $<$  0.005  & 0.002 & 0.007 & 0.001 & & \\
          & $1_2^+,0$ & & &   & $b \times 127.5$ & 127.5 & 2 & 7 & $<$  0.05  & 0.0005 & 0.004& 0.0003 & & \\
\hline
$4_1^+,0$ & $3_1^+,0$ & 1.550 & 1.577 & 1.748 & $<$  392 & 18.8 & 8 &  32 & $<$ 0.06 & 0.003 & 0.0 & 0.0 & $>$ 0.5 ps & 10.2 ps \\
          & $2_1^+,0$ &  &  &   & & 88.5 & 28 & 112 &  & &  & & & \\
\hline
$3_2^+,0$ &$1_1^+,0$ & 1.648 & 1.881 & 1.718  & $<$  43 & 33.3 & 33 & 131 &  & & &  & $>$ 1.3 ps & 1.9 ps \\
          & $3_1^+,0$ & & &   & $<$  20 & 4.8 & 2 & 7 & $<$  0.006  & 0.0 & 0.0  & 0.0 & & \\
          & $1_2^+,0$ & & &   &      & 109.4 & 4 & 18 & & & & & & \\
          & $2_1^+,0$ & & &   &      & 35.2  & 2  & 9 & & 0.002 & 0.002 & 0.0002 & & \\
          & $4_1^+,0$ & & &   &      & 20.3  & 3 & 11 & & 0.002 & 0.016 & 0.001 & & \\
\hline
$2_2^+,1$ & $1_1^+,0$ & 1.653 & 1.782 & 1.580 & $<$ 60 & 0.4 & 0.6 & 0.6 &$<$ 0.011 & 0.0005 & 1.53 & 0.82 & 50(10) fs & 30 fs \\
          & $0_1^+,1$ & & &   & 122(47) &  135.7 & 34 & 135 && & & & & \\
          & $3_1^+,0$ & & &   & $2^{+9}_{-2}$ & 1.5 & 4 & 4 & 0.57(12) &1.0 & 4.32 & 2.3 & & \\
          & $1_2^+,0$ & & &   & 27(22) & 0.5 & 2 & 2 & 0.3(1) & 0.29 & 0.23 & 0.15 & & \\
          & $2_1^+,0$ & & &   & & 1.12 & 0.2 & 0.2 & & 0.31 & 1.02  & 0.4 & & \\
          & $4_1^+,0$ & & &   & & 0.28 & 0.4 & 0.4 & & & & & & \\
          & $3_2^+,0$ & & &   & & 0.42 & 0.7 & 0.7 & & 0.065 & 0.068  & 0.04 & & \\
\hline
$5_1^+,0$ & $3_1^+,0$ & 2.066 & 1.999 & 2.578 & x & 6.8 & 0.03 & 0.1 & & & & &  & 9.6 ps \\
          & $3_2^+,0$ &       &     &  & 90(50)$\cdot$x & 90.7 & 23 & 92 & &  &  & & &\\
          & $4_1^+,0$ &       &     &  &  & 47.8 & 6 & 25 &  & 0.003 &0.0 & 0.0 & & \\
\hline
$4_2^+,0$ & $3_1^+,0$ & 2.690(20) & 2.532 &  &  & 
&  &  &  &  
&  & & &  \\
\hline
$4_3^+,1$ & $3_1^+,0$ & 2.751 & 2.682 & 2.318 &  & 0.12 & 0.6 & 0.6 & 0.02(1)$\cdot$y& 0.01 &  0.004 & 0.005 & & 70 fs \\
          & $3_2^+,0$ &  &  &          &  & 0.25 & 2.2 & 2.2& y & 0.16 & 1.30 & 0.75 & & \\
          & $2_1^+,0$ &  &  &          &  & 0.03 & 0.01  & 0.01 && &      & &   & \\
          & $2_2^+,1$ &  &  &          &  & 76.8 &  23  & 87.8  &&&      & &   & \\
          & $5_1^+,0$ &  &  &          &  & 0.0  & 0.4 & 0.4 & & 0.017& 0.001 & 0.001 & & \\
          & $4_1^+,0$ &  &  &          &  & 0.14 & 0.2 & 0.2 & 1.2(4)$\cdot$y& 0.27 & 2.75 &  1.0 & & \\ \hline
\end{tabular}
\end{center}
\end{table*}

\end{document}